\documentclass[utf8]{frontiersSCNS} 
\usepackage{url,hyperref,lineno,microtype,subcaption}
\usepackage[onehalfspacing]{setspace}

\def\keyFont{\fontsize{8}{11}\helveticabold }
\def\firstAuthorLast{Marc Wiedermann {et~al.}} 
\def\Authors{Marc Wiedermann\,$^{1,*}$, Jonatan F. Siegmund\,$^{2}$ Jonathan F. Donges\,$^{3,4}$, Reik V. Donner\,$^{2,5,6}$}
\def\Address{
$^{1}$ FutureLab on Game Theory \& Networks of Interacting Agents, Research Department IV -- Complexity Science, Potsdam Institute for Climate Impact Research (PIK) -- Member of the Leibniz Association, Potsdam, Germany \\
$^{2}$ Research Department IV -- Complexity Science, Potsdam Institute for Climate Impact Research (PIK) -- Member of the Leibniz Association,  Potsdam, Germany \\
$^{3}$ FutureLab on Earth Resilience in the Anthropocene, Research Department I -- Earth System Analysis, Potsdam Institute for Climate Impact Research (PIK) -- Member of the Leibniz Association, Potsdam, Germany\\
$^{4}$ Stockholm Resilience Centre, Stockholm University, Stockholm, Sweden\\
$^{5}$ Department of Water, Environment, Construction and Safety, Magdeburg--Stendal University of Applied Sciences, Magdeburg, Germany \\
$^{6}$ Research Department I -- Earth System Analysis, Potsdam Institute for Climate Impact Research (PIK) -- Member of the Leibniz Association, Potsdam, Germany
}
\def\corrAuthor{Marc Wiedermann, FutureLab on Game Theory \& Networks of Interacting Agents, Research Department IV -- Complexity Science,  
Potsdam Institute for Climate Impact Research (PIK) -- Member of the Leibniz Association, Telegrafenberg A31, 14473 Potsdam, Germany}

\def\corrEmail{marcwie@pik-potsdam.de}

\begin{document}
\onecolumn
\firstpage{1}

\title[Differential imprints of ENSO flavors]{Differential imprints of distinct ENSO flavors in global patterns of very low and high seasonal precipitation} 

\author[\firstAuthorLast ]{\Authors} 
\address{} 
\correspondance{} 

\extraAuth{}

\maketitle

\begin{abstract}
\noindent
  The effects of El Ni\~no's two distinct flavors, East Pacific (EP) and Central Pacific (CP)/Modoki El Ni\~no, on global climate variability have been studied intensively in recent years. Most of these studies have made use of linear multivariate statistics or composite analysis. Especially the former assumes the same type of linear statistical dependency to apply across different phases of the El Ni\~no--Southern Oscillation (ENSO), which appears not necessarily a justified assumption. Here, we statistically evaluate the likelihood of co-occurrences between very high or very low seasonal precipitation sums over vast parts of the global land surface and the presence of the respective EP and CP types of both, El Ni\~no and La Ni\~na. By employing event coincidence analysis, we uncover differential imprints of both flavors on very low and very high seasonal precipitation patterns over distinct regions across the globe, which may severely affect, among others, agricultural and biomass production or public health. We particularly find that EP periods exhibit statistically significant event coincidence rates with hydrometeorological anomalies at larger spatial scales, whereas sparser patterns emerge along with CP periods. Our statistical analysis confirms previously reported interrelations for EP periods and uncovers additional distinct regional patterns of very high/low seasonal precipitation, such as increased rainfall over Central Asia alongside CP periods that have to our knowledge not been reported so far. Our results demonstrate that a thorough distinction of El Ni\~no and La Ni\~na into their two respective flavors could be crucial for understanding the emergence of strong regional hydrometeorological anomalies and anticipating their associated ecological and socioeconomic impacts.
  
\tiny
 \keyFont{ \section{Keywords:} El Ni\~no--Southern Oscillation, Precipitation, Central Pacific El Ni\~no, Central Pacific La Ni\~na, Event Coincidence Analysis} 
\end{abstract}

\section{Introduction}
The positive (El Ni\~no) and negative (La Ni\~na) \textit{phases} of the El Ni\~no--Southern Oscillation (ENSO) are known to be associated with wide-spread anomalies in the mean hydrometeorological conditions at various distant parts of the Earth. These long-ranged interactions are often referred to as teleconnections~\citep{trenberth_definition_1997,neelin_tropical_2003, domeisen2019teleconnection}. In this context, recent findings indicate that there exist two distinct \textit{types} or \textit{flavors} of El Ni\~no phases, usually referred to as the East Pacific (EP) or canonical El Ni\~no and the Central Pacific (CP) El Ni\~no or El Ni\~no Modoki~\citep{ashok_nino_2007,kao_contrasting_2009, marathe2015revisiting}, respectively. It has been shown that these two flavors are possibly associated with distinct hydrometeorological responses in certain regions~\citep{taschetto_nino_2009}. Examples include reduced rainfall over eastern Australia~\citep{chiew_ninosouthern_1998} or Southern Africa~\citep{ratnam2014remote} only during EP periods as well as increased precipitation over the tropical regions of Africa~\citep{preethi2015impacts} or the western Indo-Pacific Oceans~\citep{weng2011anomalous,feng2014influence} during CP periods. The main reason for such differential responses are different longitudinal positions of the strongest ENSO related sea surface temperature (SST) anomalies in the tropical Pacific ocean alongside the different flavours, which results in different modifications of large scale atmospheric circulation patterns~\citep{ashok_climate_2009, domeisen2019teleconnection}.

A similar discrimination into two types has been suggested for La Ni\~na phases as well~\citep{kug_are_2011}, even though their respective imprints on SST patterns seem less distinct than for El Ni\~no~\citep{kao_contrasting_2009, ren_nino_2011}. It therefore remains an open problem to provide further statistical and/or dynamical evidence either in favor of or against such a distinction of different La Ni\~na flavors in analogy to El Ni\~no~\citep{levine2016extreme, chen2015strong, zhang2015impacts}. Still, differential hydrometeorological responses alongside specific La Ni\~na phases have recently been identified especially across the Pacific~\citep{hidayat2018impact, magee2017influence, song2017distinct, shinoda2011anomalous}, which provides good reason for discriminating La Ni\~na into two types as well.

Most previous studies (including those mentioned above) on the teleconnective impacts of different ENSO phases and flavors have either applied linear statistical tools, such as (partial) correlation analysis~\citep{diaz2001, preethi2015impacts, weng2011anomalous, hidayat2018impact, magee2017influence} or empirical orthogonal function (EOF) analysis~\citep{taschetto_nino_2009}, or investigated composites (i.e., mean spatial patterns for a specific type of ENSO period) of the corresponding climate observable of interest~\citep{hoell2014, feng2014influence, song2017distinct}. Specifically, the former methods share the common limitation of focusing on linear or average interdependencies between ENSO and possible response variables. At the same time, global climate change has been projected to lead to an increase in the strength and frequency of both, climate extremes~\citep{karl2003modern,easterling2000observed} as well as extreme ENSO phases~\citep{cai_increasing_2014, cai2015increased}. This calls for more systematically assessing possible statistical as well as dynamical/mechanistic linkages between these two findings~\citep{allan_atmospheric_2008}. Accordingly, the present work aims to identify and quantitatively characterize spatial patterns of markedly wet/dry seasons that have an elevated probability to co-occur with certain types (EP or CP) of ENSO phases, following upon previous findings that ENSO can have large-scale effects on rainfall patterns at both, global and regional scales~\citep{dai_global_2000,ropelewski_global_1987}.

Several strategies to distinguish East Pacific (EP) from Central Pacific (CP) ENSO events have been proposed in the recent past~\citep{hendon_prospects_2009}. One prominent example is the ENSO Modoki Index that is computed as the weighted average SST anomalies over three specific regions ($165^\circ$E to $140^\circ$W and $10^\circ$S to $10^\circ$N, $110^\circ$W to $70^\circ$W and  $15^\circ$S to $5^\circ$N, as well as $125^\circ$E to $145^\circ$E and $10^\circ$S to $20^\circ$N) in the equatorial Pacific~\citep{ashok_nino_2007}. Other approaches have used empirical orthogonal functions~\citep{graf_central_2012,kao_contrasting_2009} or combinations of the Nino3 and Nino4 indices~\citep{kim_unique_2011,hu_analysis_2011} to provide the desired categorization. 

\cite{wiedermann_climate_2016} recently compiled a synthesis of eight previous studies that used either of the aforementioned frameworks and identified several ENSO periods during which these techniques proposed either ambiguous, mutually inconsistent or incomplete classifications, with certain ENSO periods remaining entirely unassigned. In order to fill those gaps, \cite{wiedermann_climate_2016} proposed a new index based on spatial correlation structures among global surface air temperature anomalies, which can be conveniently studied in terms of the transitivity of so-called \textit{functional climate networks}~\citep{tsonis_what_2006,donner_complex_2017}. This transitivity index confirmed the flavors of all El Ni\~nos between 1953 and 2013 that had been classified in mutual agreement across the existing literature~\citep{wiedermann_climate_2016}. It also allowed to assign types to those cases where former work yielded incomplete or ambiguous categorizations and thereby provided a consistent, comprehensive and complete classification of the respective flavours. Moreover, \cite{wiedermann_climate_2016} showed that the transitivity index also naturally distinguishes La Ni\~na episodes into two corresponding types, thereby providing a unique advantage over other classification schemes that mainly focus on El Ni\~no phases alone. Even though the existing literature on a discrimination of different La Niña periods is comparatively scarce, the transitivity index confirmed the results of at least two recent studies~\citep{yuan2013different, tedeschi2013influences} and again provides a consistent classification for years that previously had types ambiguously or not all assigned~\citep{wiedermann_climate_2016}. 
For both reasons, i.e., the comprehensiveness of the classification and the ability to distinguish La Ni\~na into two types as well, we will directly use the classified periods from~\cite{wiedermann_climate_2016} (see Tab.~\ref{tab:enso_years} for an overview) for the purpose of our present study. 


Based on this categorization of El Ni\~no and La Ni\~na phases, we quantify the likelihood of simultaneous or time-delayed co-occurrences of strong seasonal wet/dry anomalies at a local scale across vast parts of the global land surface with a certain type of ENSO phase. Specifically, we consider seasonal precipitation sums for boreal fall (September to November, SON), winter (December to February, DJF) and spring (March to May, MAM) as key seasons of the developmental cycle of El Ni{\~{n}}o and/or La Ni{\~{n}}a conditions. By contrast, we omit the boreal summer season, even though large scale hydrometeorological conditions during this season could still be distinctively affected by different (approaching or withdrawing) ENSO phases. The reason for this choice is that there are various cases where El Ni{\~{n}}o and La Ni{\~{n}}a conditions occur in subsequent years so that a unique attribution of anomalies to any of these two phases would be hardly possible. In our present analysis, strong wet (dry) anomalies will be defined as seasonal precipitation sums exceeding the empirical 80th (falling below the empirical 20th) percentile of the distribution of all seasonal values on record for a given season and location. 

To statistically quantify co-occurrences between different types of ENSO phases and very wet/dry seasons, we employ \textit{event coincidence analysis} (ECA)~\citep{donges_nonlinear_2011,Donges2016a}. Put in simple terms, ECA counts the fraction of events of one type (here, some particularly wet/dry season at a certain location) that coincide with those of another type (here, some type of ENSO period) while, in contrast to other conceptually related approaches like event synchronization~\citep{quian_quiroga_event_2002,boers2014prediction, malik2010spatial}, allowing for a precise control of the relative timing between them~\citep{wolf2020event}. This statistical framework has already been successfully applied to quantify the likelihood of climatic (extreme) events possibly triggering certain ecological or socioeconomic responses, such as extreme annual~\citep{rammig_coincidences_2015} and daily~\citep{Siegmund2016b} tree growth or flowering dates~\citep{Siegmund2016a}, the outbreak of epidemics~\citep{Donges2016a} or armed conflicts~\citep{Schleussner16082016,Ide2020}. 

In the course of this work, we first evaluate significant event coincidence rates for the canonical (EP) El Ni\~no. This allows us to demonstrate the consistency of our approach by comparing the obtained spatial patterns with a variety of previously reported results. We continue by studying co-occurrences between strong seasonal precipitation anomalies and the so far less intensively studied CP El Ni\~no periods to highlight differences in their teleconnectivity patterns as compared to their EP counterparts. Ultimately, we also study La Ni\~na periods and demonstrate that the most remarkable large scale spatial patterns of strong seasonal rainfall anomalies nonrandomly co-occurring with this negative ENSO phase are associated with the corresponding EP flavor. This finding suggests that in the light of recent discussions on the existence of two distinct types of La Ni\~na periods~\citep{kug_are_2011, kao_contrasting_2009, ren_nino_2011, zhang2015impacts}, it is indeed meaningful to consider a global impact based distinction between one type that significantly affects seasonal wet/dry patterns globally and another that exhibits less spatially coherent impacts. Wherever appropriate, we also briefly discuss possible ecological and socioeconomic consequences of the identified seasonal precipitation anomalies. 

\section{Data \& Methods}

\subsection{GPCC rainfall data}

We utilize gridded monthly precipitation data provided by the Global Precipitation Climatology Centre (GPCC) at a spatial resolution of $2.5^\circ \times 2.5^\circ$~\citep{schneider_gpcc_2018}. Since reliable 
discriminations of El Ni\~no and La Ni\~na periods into their respective EP and CP flavours are so far mainly available for the second half of the 20th century~\citep{wiedermann_climate_2016,freund2019higher,graf_central_2012, yuan2013different}, we restrict our analysis to the period from 1951 to 2016 (the most recent year in the GPCC data set). We derive separate records for three seasons $s$ by aggregating the precipitation sums of the corresponding three-month periods from September to November (SON, 1951-2015), December to February (DJF, 1951/52-2015/16), and March to May (MAM, 1952-2016) to cover all seasons associated with the 1951/52 El Ni\~no and all following ENSO phases while ensuring the same length of all series. This results in three time series $P_{i,s}(t)$ per grid cell $i$ with $M=65$ annual values each. Note that the density of stations from which the GPCC data has been derived varies between 0 and more than 100 per grid cell and year~\citep{lorenz2012}, which generally results in a lower accuracy and reliability of the data for those areas with only few stations~\citep{rudolf_terrestrial_1994}. We therefore consider only grid cells with at least one station present for 95\% of the study period in a specific season
$s$ (SON, DJF, or MAM). In addition, we exclude those grid cells where the average precipitation sum in a specific season is below 3~cm (i.e., 1~cm per month) since this particular choice of threshold ensures the exclusion of deserts from our analysis~\citep{houston_variability_2006, thomas_twentieth-century_2018, chatterton_monthly_1971}. Both preprocessing steps yield a total number of $N_{SON}=1763$, $N_{DJF}=1610$, and $N_{MAM}=1736$ valid grid cells in SON, DJF, and MAM (see Supplementary Material Fig.~S5 for details on the spatial distribution of valid grid cells). 

We note that in the context of the present work, using three-month precipitation sums provides information fully equivalent to that obtained when using the three-month standardized precipitation index \mbox{SPI-3}~\citep{svoboda2012standardized, guttman1999accepting}, which is a solely precipitation-based characteristic commonly employed in drought-related studies. Specifically, SPI-3 values can be derived by a sophisticated monotonic transformation of the three-month precipitation sums, which does not change the rank order of the values and, hence, the timing of the events considered in this study.\\

\subsection{Classification of ENSO periods}
We use the classification of East Pacific (EP) and Central Pacific (CP) flavors of recent El Ni\~no and La Ni\~na episodes from~\cite{wiedermann_climate_2016} (Tab.~\ref{tab:enso_years}). This classification has been based on a comprehensive literature review of eight (two) studies that distinguished certain El Ni\~nos (La Ni\~nas) between 1953 and 2010 into their two respective types. Based on this compilation, ~\cite{wiedermann_climate_2016} identified eleven ENSO periods for which previous works yielded mutually consistent results and another ten periods for which (a subset) of previous studies yielded incomplete, ambiguous or mutually inconsistent classifications. 

In order to achieve a consistent and comprehensive classification of both, El Ni\~no and La Ni\~na flavors, including also the ENSO periods without previous consensus, \cite{wiedermann_climate_2016} introduced the so-called \textit{transitivity index}, which reflects distinct characteristics (in terms of abundance and localization) of ENSO's teleconnections during EP and CP periods~\citep{radebach_disentangling_2013, wiedermann_climate_2016}. In brief, this index is obtained from one-year sliding-window lag-zero absolute correlation matrices between time series at all pairs of grid cells in the global daily surface air temperature anomaly field~\citep{kistler_ncepncar_2001}. The most relevant information of these matrices is contained in the $0.5\%$ strongest absolute correlations~\citep{donges_complex_2009, radebach_disentangling_2013, wiedermann_climate_2016}, yielding thresholds for each considered time window below which the matrix coefficients are put to zero. The thus obtained sparse matrices are then considered as weighted adjacency matrices of so-called {\em functional climate networks}~\citep{donges_complex_2009, wiedermann_climate_2016, radebach_disentangling_2013} with positive entries indicating a strong statistical relationship between climate variability in two grid cells. The transitivity index \citep{newman_structure_2003, antoniou_statistical_2008} then describes the degree to which such strong relationships among triples of grid cells are transitive, i.e., the fraction of cases in which connections between two pairs $(i,j)$ and $(i,k)$ of grid cells sharing a common member are accompanied by a third connection between the remaining pair $(j,k)$. In order to provide a representative property, this fraction is further weighted by the respective cell-sizes and strengths of the involved links, i.e., the corresponding values of the absolute cross-correlation~\citep{saramaki_generalizations_2007}. The transitivity index quantifies the (dispersed or localized) spatiotemporal distribution of links in the climate network and directly reflects distinct characteristics in the temporal evolution of spatial autocorrelations and global teleconnections that are unique to specific ENSO flavours~\citep{radebach_disentangling_2013, wiedermann_climate_2016}.

Given the presence of either El Ni\~no or La Ni\~na conditions as indicated by the Oceanic Nino Index (ONI), the transitivity index then indicates EP phases by a strong peak co-occurring with the respective ENSO period. By contrast, CP phases can be identified by the absence of such a transitivity peak with values of the index close to its baseline (see \cite{wiedermann_climate_2016} for a comprehensive description and interpretation of the framework and all necessary mathematical details). 

The transitivity index confirmed the classification of all eleven consistently reported El Ni\~no periods from earlier works and additionally provided a comprehensive classification for those periods that were previously classified inconsistently or incompletely. While most other recent approaches for classifying ENSO flavors were mostly tailored to El Ni\~no events, the transitivity index also provides a consistent classification of La Ni\~na periods~\citep{wiedermann_climate_2016}, thereby making it particularly useful for our present study. 

In total,~\cite{wiedermann_climate_2016} identified six EP El Ni\~no and sixteen CP El Ni\~nos between 1951 and 2014, which is largely consistent with a recent study by~\cite{freund2019higher} that used objective supervised machine learning to obtain a similar discrimination. Similarly, \cite{wiedermann_climate_2016} found seven EP La Ni\~nas and eleven CP La Ni\~nas. Since~\cite{wiedermann_climate_2016} only provide classifications for all ENSO periods prior to the year 2014, we apply their methodology to extend the event categorization to our present study period. We specifically identify the 2014/2015 and 2015/2016 El Ni\~no periods as CP and EP types, respectively, which is again consistent with the recently proposed classification by~\cite{freund2019higher}. An overview of all ENSO periods and types that are ultimately used in this study is given in Tab.~\ref{tab:enso_years}.

\begin{figure}[h!]
  \centering
  \includegraphics[width=0.95\linewidth]{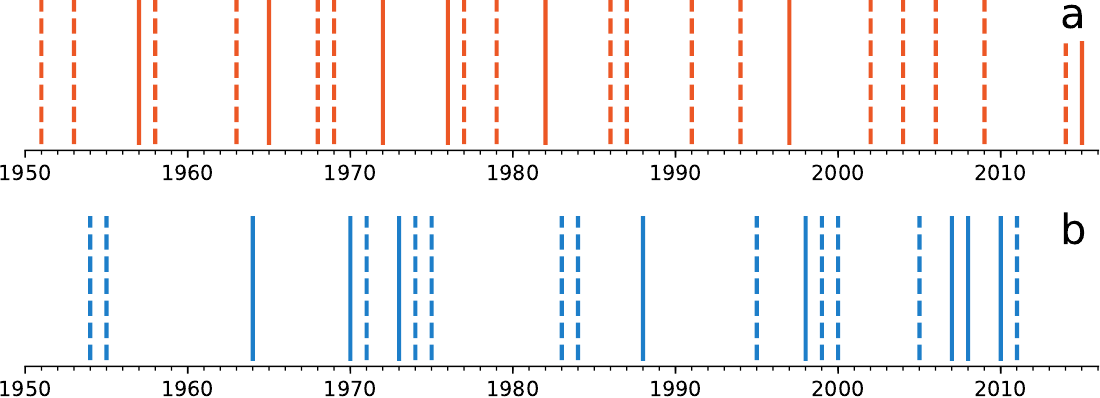}
  \caption{(a) Event series $X_\text{EPEN}$ of East Pacific (solid lines) and $X_\text{CPEN}$ of Central Pacific (dashed lines) El Ni\~nos according to the classification in~\citet{wiedermann_climate_2016} and Tab.~\ref{tab:enso_years}. (b) The same for La Ni\~na periods.
  }
 \label{fig:indices}
\end{figure}

Based on the classification in Tab.~\ref{tab:enso_years}, we create binary indicator time series for the four different ENSO phases (see Fig.~\ref{fig:indices}), e.g., an EP El Ni\~no series $X_\text{EPEN}(t)$ with $X_\text{EPEN}=1$ if $t$ marks the onset-year of an EP El Ni\~no (solid lines in Fig.~\ref{fig:indices}a). Correspondingly, we obtain the event series $X_\text{CPEN}(t)$ for CP El Ni\~nos (dashed lines in Fig.~\ref{fig:indices}a). The same procedure is applied to La Ni\~na periods, resulting in two event series $X_\text{EPLN}(t)$ and $X_\text{CPLN}(t)$, respectively (Fig.~\ref{fig:indices}b).\\\\

\subsection{Data preprocessing}\label{sec:preprocessing}

\begin{figure}[h!]
  \centering
  \includegraphics[width=0.95\linewidth]{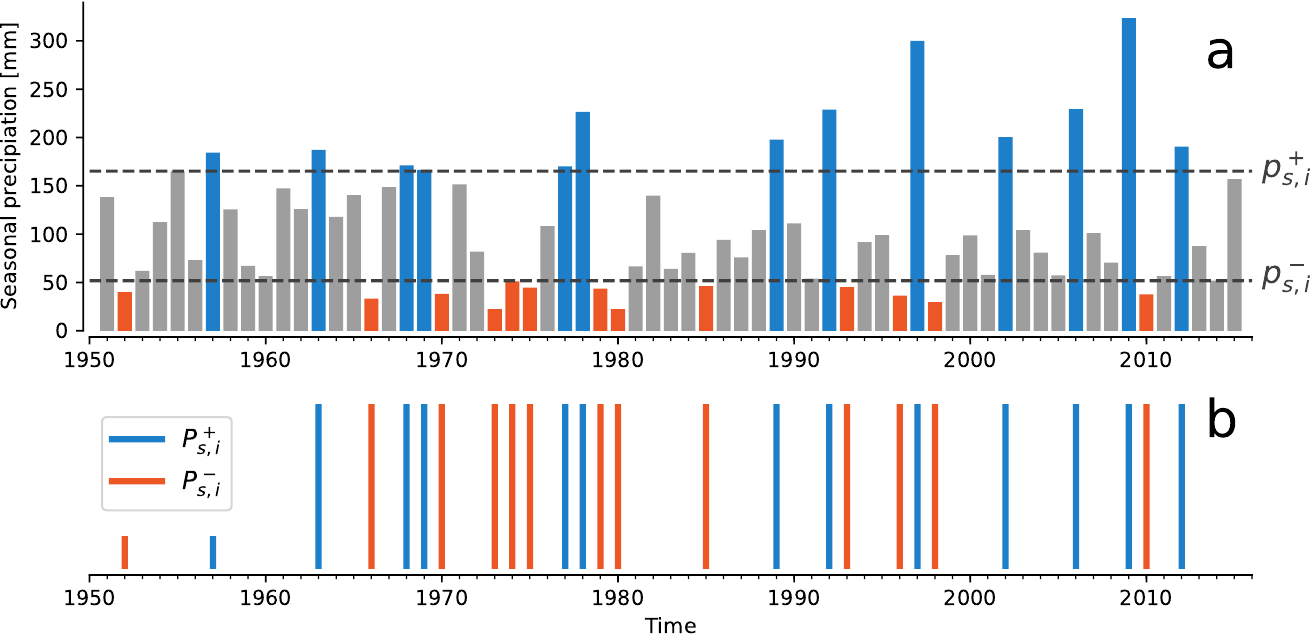}
  \caption{(a) Exemplary seasonal precipitation time series $P_{s,i}(t)$ (colored bars) as well as corresponding thresholds $p^+_{s_i}$ and $p^-_{s_i}$ computed as the 80th and 20th percentile of $P_{s,i}(t)$, respectively. Precipitation signals above $p^+_{s_i}$ (blue bars) are considered very wet while those below $p^-_{s,i}$ (red bars) are considered very dry for the subsequent analysis. (b) The two corresponding event series $P^+_{s,i}(t)$ and $P^-_{s,i}(t)$ with $P^\pm_{s,i}(t)=1$ if a corresponding event occurs at time $t$ and $P^\pm_{s,i}(t)=0$ otherwise.}
 \label{fig:rainfall_example}
\end{figure}

We identify years with seasons $s$ (DJF, SON, or MAM, see above) exhibiting extraordinary high or low precipitation amounts from the corresponding time series $P_{s,i}(t)$ for each grid cell $i$, individually. Specifically, we consider values above (below) the 80th (20th) percentile $p_{s,i}^+$ ($p_{s,i}^-$) in each of the time series $P_{s,i}(t)$ as extraordinary high (low) seasonal precipitation sums (Fig.~\ref{fig:rainfall_example}a). We choose these particular thresholds to ensure the presence of a sufficient number of particularly dry and wet seasons that is comparable with the number of different types of ENSO periods in the considered study period. It has been checked that the results obtained in this work do not change qualitatively if more restrictive or loose thresholds are applied (see Supplementary Material for details).

According to these considerations, we obtain six binary (indicator) time series $P^\pm_{s,i}(t)$ for each GPCC grid cell $i$,
\begin{equation}
  P^\pm_{s,i}(t) = \Theta(\pm P_{s,i}(t) \mp p^\pm_{s,i}),
\end{equation}
where $P^{+}_{s,i}(t)=1$ ($P^{-}_{s,i}(t)=1$) indicates the presence of a very high (very low) seasonal precipitation sum at grid cell $i$ during season $s$ in year $t$ (Fig.~\ref{fig:rainfall_example}). By following the above procedure, no further deseasonalisation of the precipitation data is necessary, since the grid cell specific seasonality of precipitation is already taken into account. Furthermore, the events are defined for each grid cell independent of all the others, which means that the specific characteristics of rainfall variability and strength at one location do not influence the definition of events at other locations as it would be the case when considering, e.g., a global (location-independent) threshold for seasonal precipitation.\\\\

\subsection{Event Coincidence Analysis}\label{sec:eca}

Event coincidence analysis (ECA) is a statistical tool that quantifies the empirical likelihood of co-occurrences of events in two series~\citep{donges_nonlinear_2011,rammig_coincidences_2015, Donges2016a}. To complement other conceptually related approaches like event synchronization~\citep{quian_quiroga_event_2002,boers2014prediction, malik2010spatial}, ECA has been developed based on analytical considerations on paired point processes and combines a precise control of the relative timing of (instantaneous or mutually lagged) events that are considered synchronous with analytical confidence bounds on the obtained event coincidence rates~\citep{donges_nonlinear_2011,Donges2016a}. Both aforementioned features provide some considerable benefit of ECA in the specific context of the present work not shared by other similar methods. Particularly the precise control of the relative timing between events (along with the specific ability to chose a global coincidence interval~\citep{wolf2020event}) is crucial for our analysis. It ensures that the we only consider precipitation anomalies that occur simultaneously (for SON or DJF) or exactly a year subsequent to the onset (for MAM) of a respective ENSO phase. 

In the context of our work, ECA provides for each grid cell the fraction (called \textit{event coincidence rate}) of EP/CP ENSO periods that co-occur with very high or low precipitation sums in SON or DJF of the same year or MAM of the year following the onset of an El Ni\~no or La Ni\~na phase. Hence, the event coincidence rate $ECR^\pm_{s,i,\bullet}$ for one pair of ENSO and precipitation event
series is given by
\begin{equation}
    ECR^\pm_{s,i,\bullet} = \frac{\sum_t X_{\bullet}(t)P^\pm_{s,i}(t - \tau)}{\sum_t
      X_\bullet(t)}.~\label{eqn:ecr}
\end{equation}
Here, $X_{\bullet}(t)$ represents one of the four time series indicating the presence of EP and CP flavors of El Ni\~no and La Ni\~na (see above). While interpreting $t$ as the calendar year, the offset $\tau$ reads $\tau=0$ for SON and DJF and $\tau=1$ for MAM. 

Note, that our present analysis studies so-called {\em trigger coincidence rates}~\citep{Donges2016a} that quantify the likelihood of a given ENSO period and phase to be {\em followed} by a specific strong/wet precipitation signal. In contrast, a complementary definition of ECR (denoted {\em precursor coincidence rate}~\citep{Donges2016a}) would address the inverse problem of quantifying likelihoods that a given precipitation event (that could arise through a variety of conditions and drivers) is {\em preceded} by a specific ENSO period and phase. However, since our present work only considers instantaneous coincidences (see~\cite{Donges2016a} for details), i.e., coincidences between events in the same ENSO period, the two types of coincidence rates only differ in their normalization (i.e., the denominator of Eq.~\eqref{eqn:ecr}). They therefore provide essentially similar information, except for possible differences in the associated statistical significance resulting from different numbers of events (i.e., different sample sizes). 

To assess the statistical significance of the empirical event coincidence rates, we assume both involved event sequences to be distributed randomly, independently and uniformly~\citep{donges_nonlinear_2011,Donges2016a}. A corresponding $p$-value is derived analytically from the probability distribution of event coincidence rates that would occur by chance only. We consider an empirical event coincidence rate as statistically significant if its associated $p$-value is smaller than a confidence level of $\alpha=0.05$~\citep{Donges2016a}.

\section{Results}\label{sec:results}

\subsection{Seasonal wet/dry patterns and EP El Ni\~no}
\label{sec:ep_el_nino}

We first investigate co-occurrences of EP El Ni\~no periods and very wet/dry seasons. Figures~\ref{fig:eca_el_nino}a,c,e highlight areas with significant event coincidence rates between EP El Ni\~nos and very dry (red squares) and wet seasons (blue squares) in SON, DJF and MAM, respectively.

During those SON seasons that correspond to the developing stages of EP El Ni\~nos, we find an elevated probability of very dry conditions over Indonesia, the Philippines and the southwestern Pacific islands as well as over northern South America and the northern Amazon Basin (Fig.~\ref{fig:eca_el_nino}a). Droughts in the latter region have been previously linked to an increased risk of biomass loss in the Amazon which normally serves as a long term carbon sink~\citep{Phillips1344, lewis_2010_2011}.

For the same season (SON), we also observe an increased likelihood of very wet conditions along the west coast of North America (Fig.~\ref{fig:eca_el_nino}a). Similarly, unusually wet conditions frequently emerge over Ecuador and southeastern South America in SON and DJF (Fig.~\ref{fig:eca_el_nino}a,c). We further observe more spatially confined regions with wet conditions over parts of the Chilean Andes in SON (Fig.~\ref{fig:eca_el_nino}a), which may result in an increased risk for the occurrence of floods in this area~\citep{bookhagen2012spatiotemporal,boers2014prediction}. All these observations agree well with previous studies~\citep{diaz2001}.

Coinciding with EP El Ni\~nos, we also observe more frequent wet conditions over the Mediterranean region and East Africa during SON (Fig.~\ref{fig:eca_el_nino}a), both of which have been previously reported in local case studies~\citep{shaman2010, Camberlin2001}. The observed tendency towards very dry conditions in southwestern Africa during DJF (Fig.~\ref{fig:eca_el_nino}c) has also been reported recently~\citep{hoell2014}. 

For MAM seasons, we observe pronounced large-scale patterns of significant event coincidence rates between EP El Ni\~nos and low precipitation sums in Northeast Brazil (Fig.~\ref{fig:eca_el_nino}e), which is consistent with previous studies~\citep{kane1997prediction}. Furthermore, the dry conditions over the Philippines that are observed during SON (see above) and DJF also persist into the MAM season (Fig.~\ref{fig:eca_el_nino}e). In addition, strong MAM rainfall occurring alongside EP El Ni\~nos is most prominently observed in the southeastern United States, which is again consistent with previous works that used composite analysis to determine North American weather patterns associated with El Ni\~no conditions~\citep{ropelewski1986north}. 

Taken together, the results of ECA are overall in good agreement with previously reported interrelations between El Ni\~no and global precipitation patterns, which have mostly been identified using linear statistical tools such as correlation analysis~\citep{Phillips1344, Camberlin2001}, principal component analysis~\citep{diaz2001} or composites based on seasonal averages~\citep{shaman2010, hoell2014}. Thus, we conclude that the application of ECA to unveil strong ENSO related hydrometeorological anomalies provides consistent results when compared to those of previous studies. Our results also imply that strong responses of seasonal precipitation to canonical (EP) El Ni\~no conditions show similar spatial patterns as the average statistical interdependency between ENSO related indices and the corresponding hydrometeorological observables. However, we note that most previous studies have not discriminated between the two El Ni\~no flavors. Thus, the agreement between our results for EP El Ni\~nos and the existing literature suggests that the previously observed (linear) effects might be dominated by the (on average stronger~\citep{kug_two_2009, huang2016ranking}) EP events.\\

\subsection{Seasonal wet/dry patterns and CP El Ni\~no}

\begin{figure*}[h!]
  \centering
  \includegraphics[width=\linewidth]{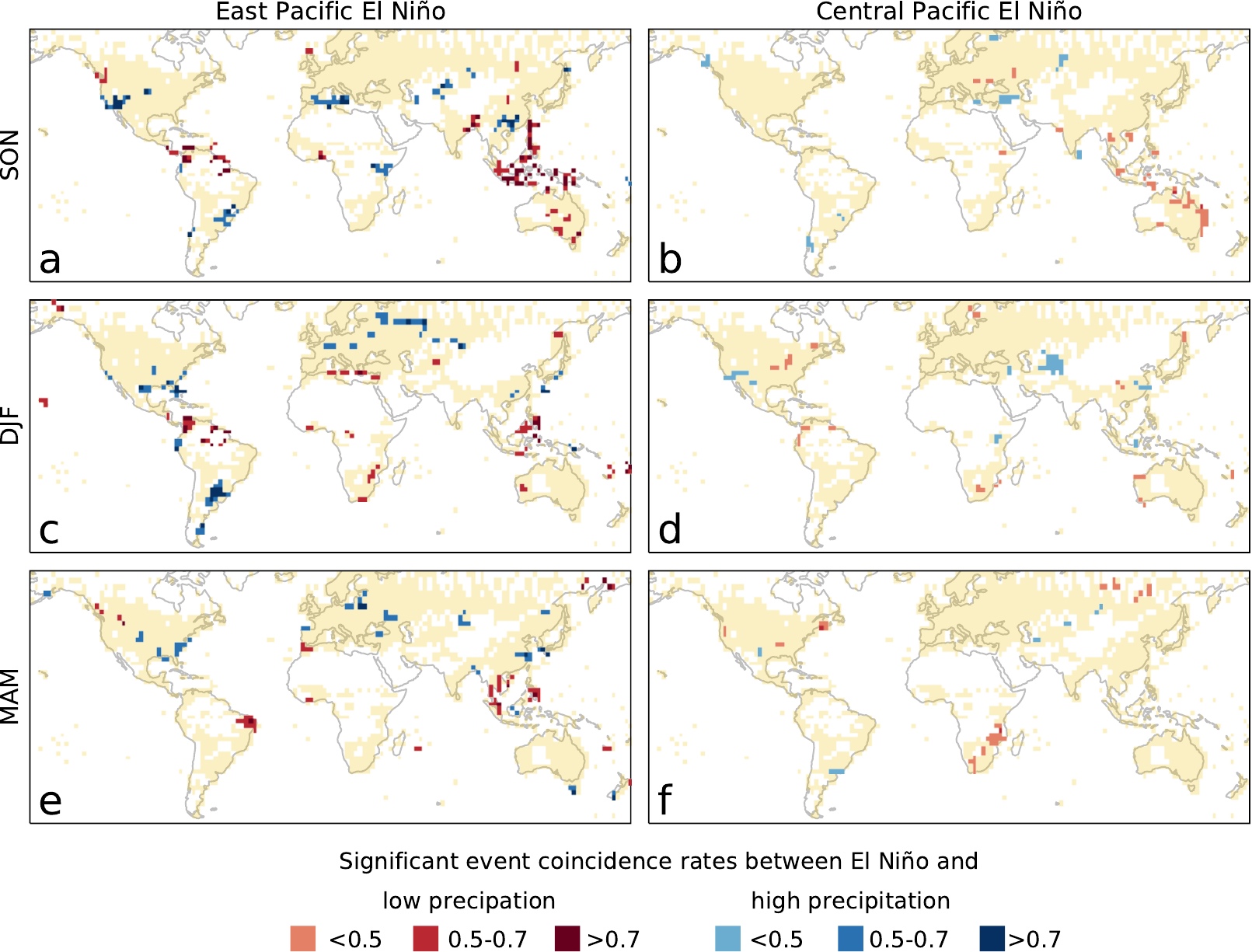}
  \caption{Statistically significant event coincidence rates (ECR) between EP (left column) and CP (right column) El Ni\~nos and very dry (red squares) or very wet (blue squares) conditions for the three seasons SON, DJF, and MAM. Dry/wet periods are defined by seasonal precipitation sums below/above the 20th/80th percentile of all years from 1951 to 2016. Only significant clusters of at least two adjacent grid-cells are shown. Different shades of red/blue indicate increasing magnitudes of significant ECR between the respective El Ni\~no events and strong/weak seasonal precipitation. Yellow areas indicate grid cells with non-significant event coincidence rates. White areas over land indicate insufficient quality of the GPCC dataset (i.e., grid cells excluded from our analysis).}
  \label{fig:eca_el_nino}
\end{figure*}

Next, we focus on very high and low seasonal precipitation along with the so far less intensively studied CP El Ni\~no (cf.\, Fig.~\ref{fig:eca_el_nino}b,d,f). We first discuss those regions that display significant event coincidence rates for EP El Ni\~nos (see above) but \textit{not} for the corresponding CP periods. Notably, very dry seasons over tropical South America that frequently co-occur together with EP El Ni\~nos are markedly less prominent for CP El Ni\~nos during DJF (Fig.~\ref{fig:eca_el_nino}d) and do not display any significant event coincidence rates at all in SON (Fig.~\ref{fig:eca_el_nino}b). The latter also holds true for very wet seasons along the western coast of Central and North America that have been observed for EP El Ni\~nos. In the same manner, the wet SON patterns over southern China, the Mediterranean, and East Africa frequently co-occurring with EP El Ni\~nos in SON do not exhibit significant coincidence rates with CP El Ni\~nos (Fig.~\ref{fig:eca_el_nino}b). We further note, that the large scale dry events over Indonesia observed along with EP El Ni\~nos during SON become less spatially coherent for CP El Ni\~nos (Fig.~\ref{fig:eca_el_nino}b). For MAM, the wet patterns over the northern Iberian peninsula that are observed for EP El Ni\~nos cannot be identified for CP El Ni\~nos (Fig.~\ref{fig:eca_el_nino}f). 

While the aforementioned observations indicate decreased or weakened impacts of CP El Ni\~nos in comparison with the EP flavor, we also observe new additional patterns of significant event coincidence rates that are not present during EP El Ni\~nos but emerge only along with CP periods. Most notably, very dry conditions become more likely along Australia's east coast during SON (Fig.~\ref{fig:eca_el_nino}b). Such hydrometeorological anomalies could thus result in severe impacts on river ecosystems and agriculture in that region~\citep{leblanc2012review}. In particular, marked drought phases in Eastern Australia are likely to cause a cascade of low river inflows, general water scarcity and large scale floodplain forest mortality as well as an increase of toxicity in the surrounding lakes~\citep{leblanc2012review}. These natural hazards can have substantial effects on agricultural production in terms of a severe reduction in irrigated crop yields~\citep{vandijk2013}. 

In addition to reduced rainfall responses, significant event coincidence rates with very wet conditions are found over southern Chile pointing towards increased rainfall during CP El Ni\~nos as compared to their canonical counterparts. In DJF months coinciding with CP El Ni\~no periods, we observe the emergence of new wet patterns over Central Asia well as a dry pattern over the north of Peru and Ecuador (Fig.~\ref{fig:eca_el_nino}d). Finally, we observe a pronounced dry pattern over Southeast Africa in MAM (Fig.~\ref{fig:eca_el_nino}f). 

Generally, we note lower event coincidence rates between seasonal wet/dry conditions and CP El Ni\~nos as compared to EP periods (cf.\ Fig.~\ref{fig:eca_el_nino}a,c,e and Fig.~\ref{fig:eca_el_nino}b,d,f). 
This might be partly explained by the larger of number of 17 CP events as compared to 7 EP events over the considered study period which is consistent with previously reported frequencies of the two flavours~\citep{hendon_prospects_2009,graf_central_2012, preethi2015impacts, wiedermann_climate_2016}. In addition, recent studies suggest that CP El Ni\~nos might be further discriminated into two subtypes based on their specific impacts on Pacific rainfall and the modes of the Indian Ocean dipole~\citep{wang2013classifying, wang2014different, wang2018new}. Along those lines, future work should therefore investigate whether the comparatively lower significant event coincidence rates between CP El Ni\~nos and strong/weak seasonal precipitation can indeed also be attributed to the presence or absence of any of these two possible subtypes.\\

\subsection{Seasonal wet/dry patterns and EP/CP La Ni\~na}

\begin{figure*}[ht!]
  \centering
  \includegraphics[width=\linewidth]{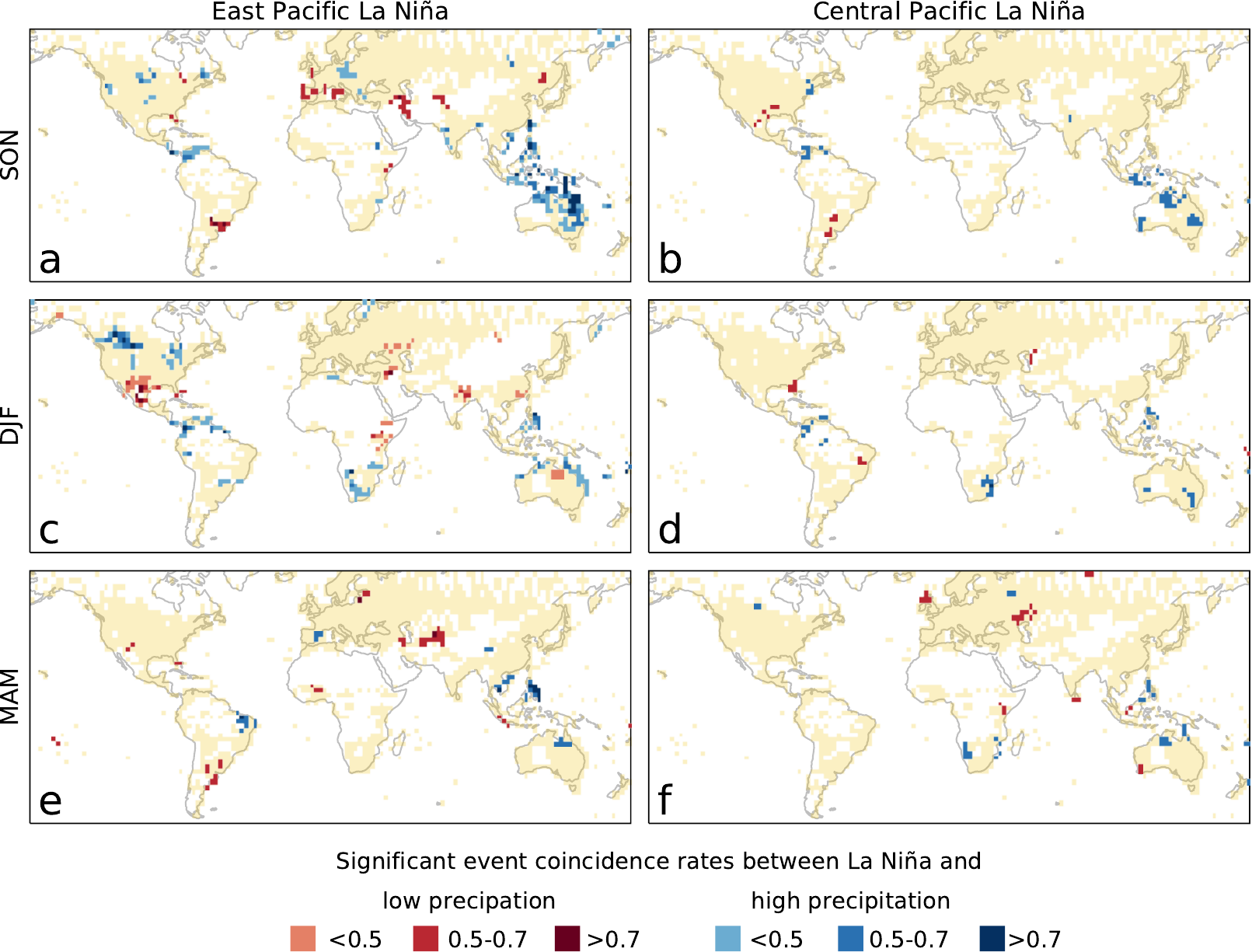}
  \caption{Same as Fig.~\ref{fig:eca_el_nino}, but for the two types of La Ni\~na.
  }\label{fig:eca_la_nina}
\end{figure*}

Ultimately, we perform the same analysis as above for La Ni\~na periods. For the EP, i.e. canonical, La Ni\~na phases (Fig.~\ref{fig:eca_la_nina}a,c,e) we again find various patterns that have already been reported in previous studies. Specifically, during SON coinciding with EP La Ni\~nas (Fig.~\ref{fig:eca_la_nina}a), we recapture an increased probability for very wet conditions over Australia and Indonesia~\citep{arblaster_interdecadal_2002} and exceptionally dry conditions in southern Europe~\citep{pozo-vazquez_ninosouthern_2005} and the south of Brazil and Uruguay~\citep{ropelewski_quantifying_1996}. We further observe significant event coincidence rates for dry seasons in the Middle East contrasted by more intense than normal rainfall over central Europe. For DJF, our analysis confirms previous findings of very wet conditions over South Africa and dry episodes over West Africa~\citep{nicholson_influence_2000} (Fig.~\ref{fig:eca_la_nina}c), the latter having been previously linked to potential agricultural losses~\citep{Karpouzoglou_2014} and considerable health risks in that specific area~\citep{Rataj_2016}. We further observe a prominent seasonal precipitation dipole with dry conditions over Mexico and elevated rainfall over southwestern Canada. The latter has become an important aspect of local water resource management~\citep{lute_rolw_2014}, but together with increasing air temperature and more frequent storms also poses the threat of landslides in corresponding coastal areas~\citep{Guthrie_2010}. For MAM seasons associated with EP La Ni\~nas, we observe a tendency towards strong rainfall over the Amazon~\citep{rogers_precipitation_1988} and parts of Northern Australia~\citep{arblaster_interdecadal_2002} and the Philippines (Fig.~\ref{fig:eca_la_nina}e).  

In contrast to the wide-spread spatially coherent wet/dry anomaly patterns observed for EP La Ni\~na periods, we find much fewer spatially extended structures along with CP La Ni\~nas (Fig.~\ref{fig:eca_la_nina}b,d,f). Most prominently, we recover previously reported wet conditions over parts of Australia in SON~\citep{arblaster_interdecadal_2002, cai_ninmodoki_2009} (Fig.~\ref{fig:eca_la_nina}b). Additionally, we uncover strongly reduced rainfall over Florida in DJF (Fig.~\ref{fig:eca_la_nina}d), and over the United Kingdom, Ireland and the west of Kazakhstan in MAM (Fig.~\ref{fig:eca_la_nina}f). In summary, we observe that CP La Ni\~nas co-occur with less spatially coherent precipitation responses as compared to their canonical counterparts. We also note that event coincidence rates between La Ni\~na and seasonal wet/dry precipitation signals are quantitatively more similar across EP and CP periods than observed for El Ni\~no (cf.\ Fig.~\ref{fig:eca_la_nina}a,c,e and Fig.~\ref{fig:eca_la_nina}b,d,f), which could be partly explained by the comparatively lower difference between the total number of 8 EP and 12 CP La Ni\~na events. In addition, future work should again investigate the potential for further discriminating CP La Ni\~nas into two distinct subtypes in analogy to El Ni\~no~\citep{wang2013classifying, wang2014different, wang2018new}. Such an approach would possibly allow us to attribute significant event coincidence rates between CP La Ni\~na and strong/weak seasonal precipitation signals to the presence or absence of a certain event type and help explain the different magnitude in significant event coincidence rates observed in Fig.~\ref{fig:eca_la_nina}.

\section{Discussion and Conclusions}\label{sec:conclusion}

We have carried out a detailed analysis of ENSO imprints in global patterns of very wet/dry seasons over land. Specifically, we distinguished both, El Ni\~no and La Ni\~na, into two distinct flavors (East Pacific and Central Pacific) by utilizing a classification based on an extensive literature review paired with the assessment of a recently proposed complex network-based index~\citep{wiedermann_climate_2016}. From this classification, we obtained event series describing the occurrence-times of the four distinct types of ENSO. Strong seasonal precipitation anomalies have been obtained from the globally gridded GPCC rainfall data set by identifying seasons with precipitation sums above the empirical 80th (below the empirical 20th) percentile of all values for a given grid point as very wet (very dry) periods. This definition follows the spirit of the 3-month aggregate standardized precipitation index (SPI-3), which just provides a monotonic rescaling of seasonal precipitation sums according to the local distributional characteristics of seasonal rainfall. Accordingly, our results can be interpreted in terms of seasonal drought characteristics. Modifying the considered thresholds for defining very wet/dry seasons within reasonable ranges did not qualitatively alter the results reported in this work (see Supplementary Material).

\begin{figure*}[t]
  \centering
  \includegraphics[width=\linewidth]{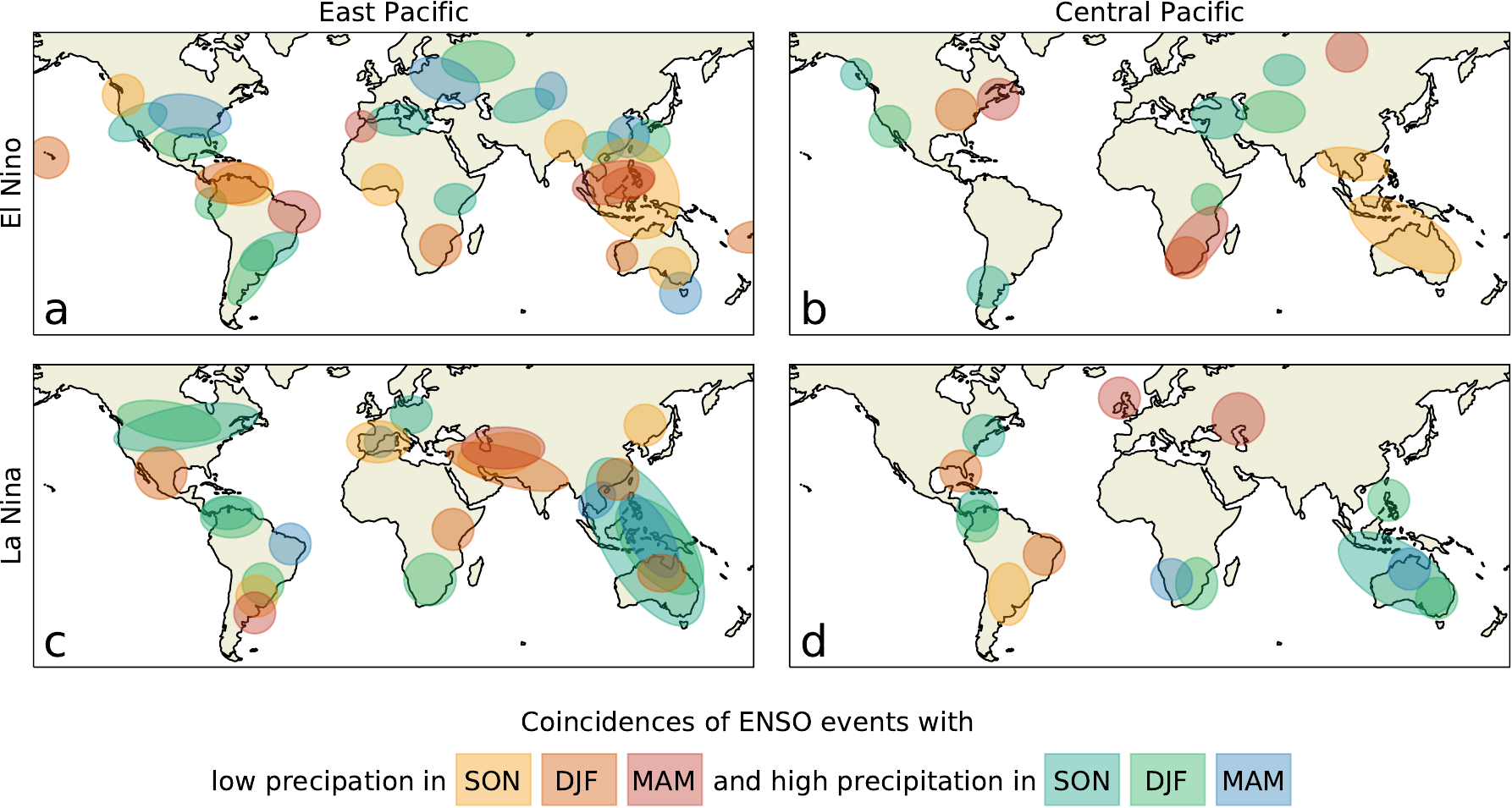}
  \caption{Schematic summary of the results presented in this work. Shaded areas indicate
    major regions in which very high or low seasonal precipitation
    sums show significant event coincidence rates with EP (a,c) or CP (b,d) El
    Ni\~no (a,b) or La Ni\~na phases (c,d). 
    }\label{fig:summary}
\end{figure*}

We have then employed event coincidence analysis~\citep{Siegmund2016b,Donges2016a} to identify grid points with significant event coincidence rates between different types of ENSO phases and very high or low seasonal precipitation sums. Our analysis confirmed that previously observed interrelationships based on linear correlation or composite analysis in many cases also apply to the timing of events corresponding to the tails of the probability distributions of seasonal precipitation sums. In addition, we identified further patterns of very wet/dry conditions with elevated probabilities alongside different ENSO types, which have to our best knowledge not been described so far. These include increased rainfall over Central Asia during CP El Ni\~nos contrasted by rainfall reduction over the same area during CP La Ni\~nas (compare also Fig.~\ref{fig:summary}b and d) which implies that for such cases standard (linear) statistical methods seem to be insufficient to unveil the underlying interrelations of events. Moreover, our present analysis demonstrates that even though a general linear relationship between ENSO and precipitation might be relatively weak or even absent for a given region, dry periods or very wet seasons can still be possible consequences of the presence of certain ENSO phases. At the same time we observe that several previously reported links between ENSO and global precipitation patterns mostly apply to East Pacific ENSO flavors as far as particularly wet/dry seasons are concerned. 

We found that the Central Pacific flavor of El Ni\~no (and in parts also that of La Ni\~na) significantly, i.e., non-randomly, co-occurs alongside wet/dry seasonal precipitation signals at lower event coincidence rates which might in parts be explained by its more frequent occurrence~\citep{hendon_prospects_2009,graf_central_2012, preethi2015impacts, wiedermann_climate_2016} compared to its East Pacific counterpart. This finding also suggests that the absolute number of event coincidences (i.e, the numerator in Eq.~\eqref{eqn:ecr}) between EP or CP ENSO periods with wet/dry seasonal precipitation per grid-cell might be of rather comparable size over the period of study, an effect that should be investigated more thoroughly in future research. In addition, recent studies suggest that the Central Pacific El Ni\~no can be further discriminated into two types with distinct imprints on precipitation signals especially in the Indo-Pacific area~\citep{wang2013classifying, wang2014different, wang2018new}. Future work should hence investigate if the comparatively lower significant event coincidence rates between CP El Ni\~no and wet/dry seasonal conditions can potentially be attributed to the presence or absence of a specific subtype of CP El Ni\~no. Since El Ni\~no and La Ni\~na show a large degree of symmetry in terms of their frequency and potential for discrimination into different types~\citep{kao_contrasting_2009, ashok_nino_2007,hidayat2018impact}, a similar analysis could also be performed for La Ni\~na as we found lower, yet significant, event coincidence rates for the CP flavor of ENSO's negative phase as well. 

Along those lines, our analysis provides a complementary impact oriented view on the recently raised question whether it is actually (statistically and/or dynamically) meaningful to distinguish La Ni\~na into two types in a similar fashion as El Ni\~no~\citep{kug_are_2011}. While some previous studies advocated for such a distinction~\citep{kao_contrasting_2009,ashok_climate_2009}, others argued that based on correlation analyses between La Ni\~na related SST patterns there is a lack of evidence for the existence of two distinct types~\citep{kug_two_2009,ren_nino_2011}. Contributing to this discussion, \cite{wiedermann_climate_2016} already demonstrated that according to the transitivity index of ENSO's global teleconnections it is indeed meaningful to provide a discrimination of La Ni\~na periods into two flavors in close analogy to El Ni\~no. Complementing recent findings on  differential hydrometeorological conditions alongside potentially distinct La Ni\~na flavors~\citep{hidayat2018impact, magee2017influence, song2017distinct, shinoda2011anomalous}, our results further demonstrate that seasonal wet/dry patterns accompanied by EP La Ni\~nas are generally more likely to arise in a spatially coherent way than such observed along with CP periods (Fig.~\ref{fig:eca_la_nina}). The same finding also applies to El Ni\~no periods, which highlights that there exists a certain symmetry between both types of ENSO phases not only in the spatial SST anomaly patterns of El Ni\~no and La Ni\~na themselves, but also with respect to their effects on global precipitation patterns. Thus, from an impact oriented point of view, our work provides further evidence in favor of a distinction between two flavors of La Ni\~na indicated by the presence or absence of very wet/dry regional conditions along with either of the two types of ENSO conditions. In other words, from the viewpoint of event based statistics (and therefore not necessarily in agreement with results based on linear correlations) it appears not only reasonable, but actually relevant to discriminate La Ni\~na into two types in a similar way as common for El Ni\~no periods.

In conclusion, our analysis provides a detailed and global overview on the large scale differential imprints of different ENSO phases and flavors in the emergence of very wet/dry seasons. All findings reported in this work are ultimately summarized in Fig.~\ref{fig:summary}, which highlights schematically the main regions where the four different types of ENSO show large scale patterns of significant event coincidence rates with very high/low seasonal precipitation. Especially with respect to the CP flavors we find a variety of regions (such as (northern) Australia or southern Africa) where El Ni\~no and La Ni\~na show opposite impacts in terms of wet/dry seasons (Fig.~\ref{fig:summary}b,d). In addition, we observe that both flavors of one ENSO phase can also display similar impacts over the same regions, such as reduced precipitation over Australia or south-east Asia for EP {\em{and}} CP El Ni\~nos (Fig.~\ref{fig:summary}a,b) or enhanced precipitation over Australia or northern South America for EP {\em{and}} CP La Ni\~nas (Fig.~\ref{fig:summary}c,d). We further uncover unique signatures that are only observed for a single type of ENSO phase, e.g, reduced precipitation over northern India and the Middle East for EP (and {\em{not}} CP) La Ni\~nas (Fig.~\ref{fig:summary}c). The general trend towards less significant clusters of large-scale coincidences between ENSO and very wet/dry seasons along with CP as compared with EP periods is observable for both, El Ni\~no and La Ni\~na, as well (c.f.\ Fig.\ref{fig:summary}a,b and Fig.\ref{fig:summary}c,d). Our results thus demonstrate that a thorough discrimination of ENSO can be crucial for properly anticipating strong regional and seasonal hydrometeorological anomalies since its differential impacts may not only vary depending on the presence of El Ni\~no and La Ni\~na conditions but are additionally modulated by their respective EP and CP flavors.

\section{Perspectives and Outlook}

Future work should further apply the statistical concepts used in this work to also study ENSO related imprints on other climate variables (e.g., surface air temperature) as well as corresponding effects on socioeconomically and ecologically relevant observables like agricultural yields or water availability. This also implies that if the specific flavor of a developing El Ni\~no or La Ni\~na was to be detected early enough, possible threats like droughts or elevated flood risks, as well as their ecological and socio-economic consequences, could be better anticipated. 

Since reliable predictions of EP or CP types of ENSO are scarce, the framework surrounding our present analysis has the potential to serve as an indirect early classification scheme of such periods based on observed globally distributed impacts. Specifically, the observation of early strong/weak climate signals either in SON or the presently ignored boreal summer (June to August, JJA) season of the onset year could be employed to assess the likelihood of approaching an ENSO phase with a specific flavor. Here, ECA could be used to analyze probabilities for specific types of seasonal climate anomalies -- that could arise through a variety of conditions and drivers -- to be {\em preceded} by certain ENSO periods and flavours (using so-called {\em precursor event coincidence rates}~\citep{Donges2016a}, see Sec.~\ref{sec:eca}). Such assessments would complement the estimation of {\em trigger coincidence rates}~\citep{Donges2016a} at which a specific ENSO flavour is {\em followed} by a certain precipitation response as it is studied in our present work. The corresponding impact based classification of ENSO periods could then be used to systematically estimate probabilities of potentially upcoming later (DJF, MAM or even JJA of the year following the onset of an ENSO period) strong/weak precipitation seasons. In this context, we recall that for the case of instantaneous co-occurrences as studied here, precursor and trigger event coincidence rates solely differ by which of the two series (ENSO or precipitation events) is considered as the reference, i.e., the denominator in Eq.~\eqref{eqn:ecr}. Since (sea or air) temperature anomalies are known to exhibit considerable persistence, it may therefore be another relevant extension of the present work to study non-instantaneous (i.e., multi-year) statistical linkages between ENSO episodes and wet/dry patterns, in which case trigger and precursor event coincidence rates become distinct statistical concepts. 

Taken together, a corresponding analysis for early classification of ENSO flavours would need to be accompanied by a thorough review of recent approaches for ENSO prediction and attribution, a proper selection of climate variables beyond precipitation that potentially exhibit a mechanistic link with ENSO variability \citep{DiCapua2019}, and an associated estimation of appropriate temporal scales. Ultimately, such a predictive analysis would require a reliable validation of the obtained results by means of dividing the considered study period into training and test intervals, potentially accompanied by extending the analysis at least over the complete 20th century or even beyond to ensure a sufficient number of events. Even though such an endeavor is clearly beyond the scope of the present work, it remains as a proposal for a potentially important subject of future research with the potential for an early attribution of EP and CP flavours to developing El Ni\~nos and La Ni\~nas.

Finally, a more detailed intercomparison between the results obtained from traditional linear and complementary event based statistics could prove useful in assessing which regions are most affected in terms of strong climate responses to the presence of any combination of ENSO phase and flavor. Specifically, and in order to not only assess timing and regional extent of coincidences between ENSO events and precipitation signals, future work should systematically combine ECA with complementary techniques such as composite analysis to also study quantitatively the (relative) changes in magnitude of enhanced/reduced precipitation occurring alongside EP or CP type phases of ENSO. While the current work has solely focused on observational data, it should also be investigated if the reported co-occurrence patterns can be equally observed in historical simulations of state of the art coupled climate models (e.g., those contributing to the Coupled Model Intercomparison Project Phase 6 (CMIP 6)), with any difference pointing to potentially insufficiently represented key processes in the models, thereby contributing to the identification of such processes and, hence, future model improvements. Ultimately, given that the frequency and magnitude of different ENSO phases might be markedly affected by global climate change~\citep{yeh2007enso, stevenson2012significant, cai_increasing_2014, cai2015increased}, it appears additionally useful to further apply the presented framework to future climate projections in order to assess possible changes in the spatial extent and frequency of ENSO related extreme events.

\section*{Conflict of Interest Statement}
The authors declare that the research was conducted in the absence of any commercial or financial relationships that could be construed as a potential conflict of interest.

\section*{Author Contributions}
All authors designed the study. M.W. and J.F.S. analysed the data and wrote the manuscript with input from all authors. J.F.D. assisted with the analysis. R.V.D. supervised the study. M.W., J.F.D. and R.V.D. substantively revised the work.

\section*{Funding}
M.W., J.F.S. and R.V.D. acknowledge funding by the German Federal Ministry for Education and Research via the BMBF projects CoSy-CC$^2$ (grant no.~01LN1306A), GOTHAM (grant no.~01LP16MA) and ROADMAP (grant no.~01LP2002B). J.F.D. has been supported by the Stordalen Foundation. M.W. and J.F.D. thank the Leibniz Society (project DOMINOES) for financial support. R.V.D. acknowledges the IRTG 1740 funded by DFG and FAPESP. J.F.S. has been supported by the Evangelisches Studienwerk Villigst e.V. 

\section*{Acknowledgements}
This paper was developed within the scope of the IRTG 1740 / TRP 2015/50122-0, funded by the DFG / FAPESP. The authors gratefully acknowledge the European Regional Development Fund (ERDF), the German Federal Ministry of Education and Research and the Land Brandenburg for supporting this project by providing resources on the high performance computer system at the Potsdam Institute for Climate Impact Research. This manuscript has been released as a pre-print at \url{http://arxiv.org/abs/1702.00218}~\citep{wiedermann2020differential}. The content of this manuscript has been partially published as part of the  dissertation of Marc Wiedermann~\citep{wiedermann2018classification}.

\bibliographystyle{frontiersinSCNS_ENG_HUMS} 
\bibliography{main.bib}

\newpage

\section*{Tables \& Table captions}

\begin{table}[h!]
\begin{tabular}{ p{2.5cm}  p{2.5cm}  p{2.5cm}  p{2.5cm}  }
 \hline
 \multicolumn{2}{c}{El Ni\~no} & \multicolumn{2}{c}{La Ni\~na} \\
 \hline\hline
 East Pacific & Central Pacific & East Pacific & Central Pacific\\
 \hline
 1957 & 1951 & 1964 & 1954\\
 1965 & 1953 & 1970 & 1955\\
 1972 & 1958 & 1973 & 1967\\
 1976 & 1963 & 1988 & 1971\\
 1982 & 1968 & 1998 & 1974\\
 1997 & 1969 & 2007 & 1975\\
 2015* & 1977 & 2010 & 1984\\
 & 1979 &  & 1995\\
 & 1986 &  & 2000\\
 & 1987 &  & 2001\\
 & 1991 &  & 2011\\
 & 1994 &  & \\
 & 2002 &  & \\
 & 2004 &  & \\
 & 2006 &  & \\
 & 2009 &  & \\
 & 2014* &  & \\
\hline
\end{tabular}
\caption{Classification of ENSO periods according to \cite{wiedermann_climate_2016} on the basis of a comprehensive literature synthesis and a consistent classification using the network-based transitivity index. The given years correspond to the onset-year of each ENSO event such that, e.g., 1951 indicates the 1951/1952 El Ni\~no event. Years marked with an asterisk are classified using the same methodology as in \cite{wiedermann_climate_2016} to extend the data to the period of study considered in this work.}
\label{tab:enso_years}
\end{table}

\newpage

\section*{Figures \& Figure captions}

\newpage

\end{document}


\onecolumn
\firstpage{1}

\title[Supplementary Material]{{\helveticaitalic{Supplementary Material}}}

\maketitle

\section{Event coincidence rates for different thresholds of strong and weak seasonal precipitation}
 
\begin{figure*}[t]
  \centering
  \includegraphics[width=\linewidth]{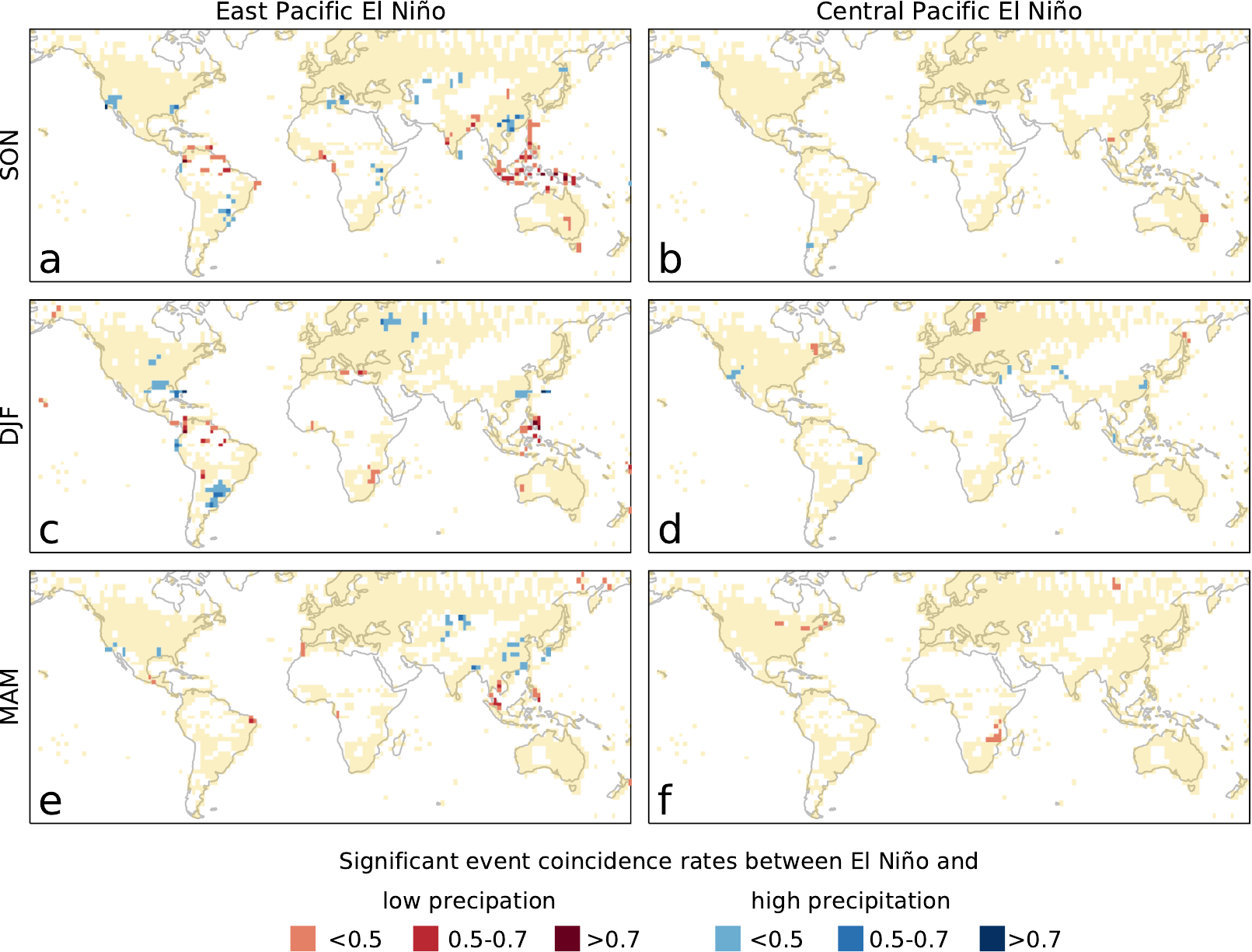}
  \caption{Same as Fig.\ 3 of the main manuscript but for strong (weak) precipitation
    events defined as values that exceed (fall below) the 90th (10th) percentile of each
    individual time series.}    
  \label{fig:eca_el_nino90}
\end{figure*}

\begin{figure*}[t]
  \centering
  \includegraphics[width=\linewidth]{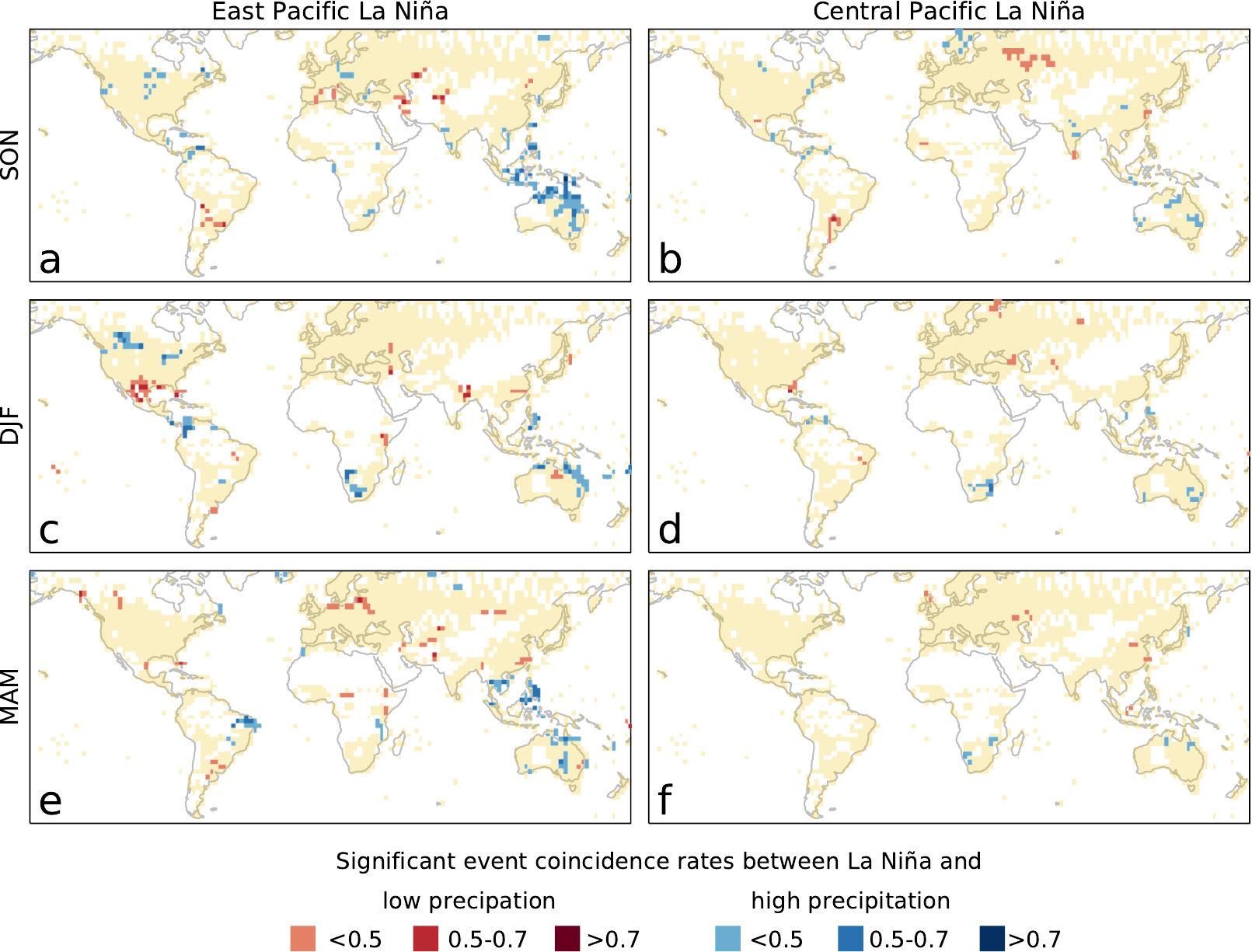}
  \caption{Same as Fig.~\ref{fig:eca_la_nina90} for La Ni\~na periods.}
  \label{fig:eca_la_nina90}
\end{figure*}

\begin{figure*}[t]
  \centering
  \includegraphics[width=\linewidth]{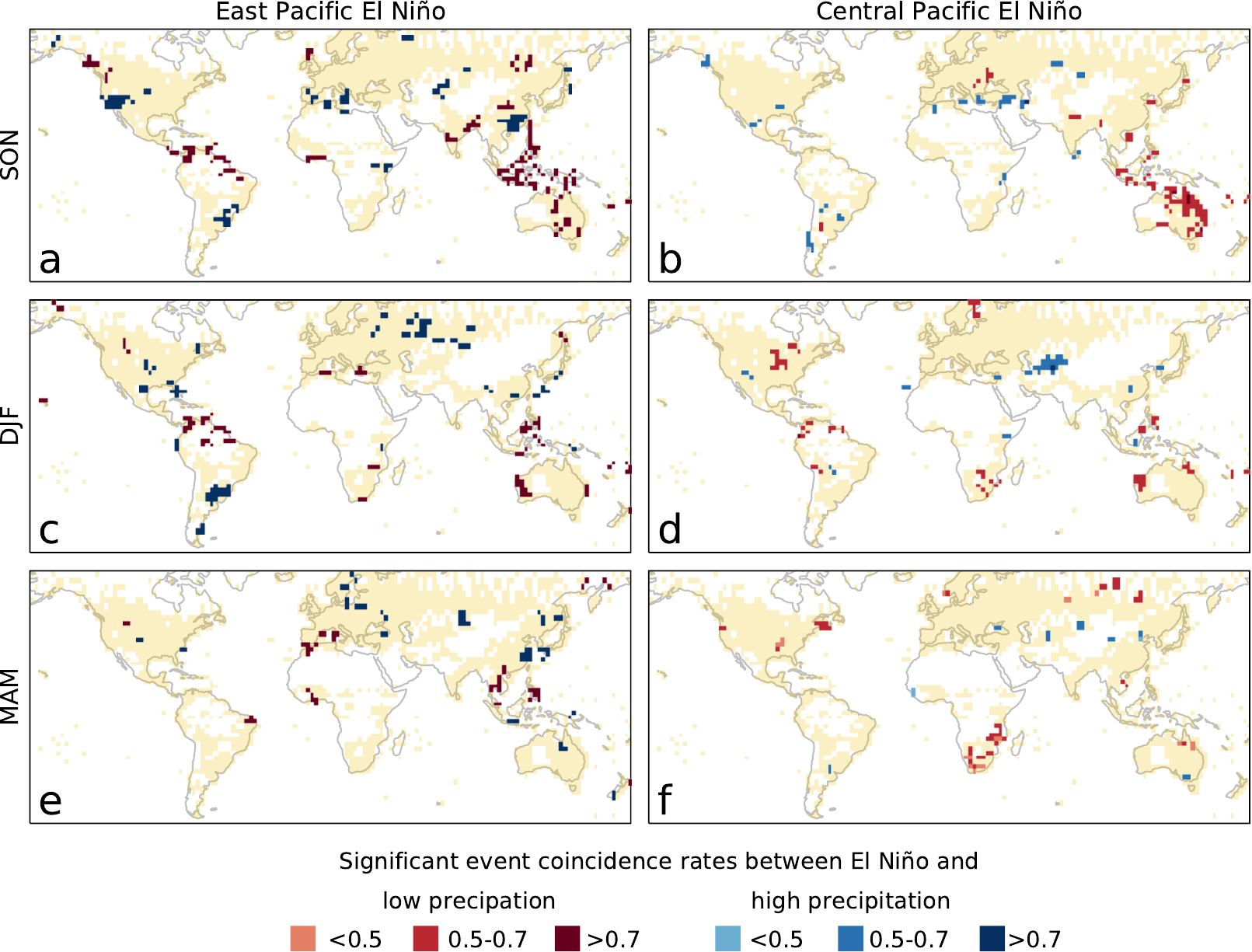}
  \caption{Same as Fig.\ 3 of the main manuscript but for strong (weak) precipitation
    events defined as values that exceed (fall below) the 70th (30th) percentile of each
    individual time series.}    
  \label{fig:eca_el_nino70}
\end{figure*}

\begin{figure*}[t]
  \centering
  \includegraphics[width=\linewidth]{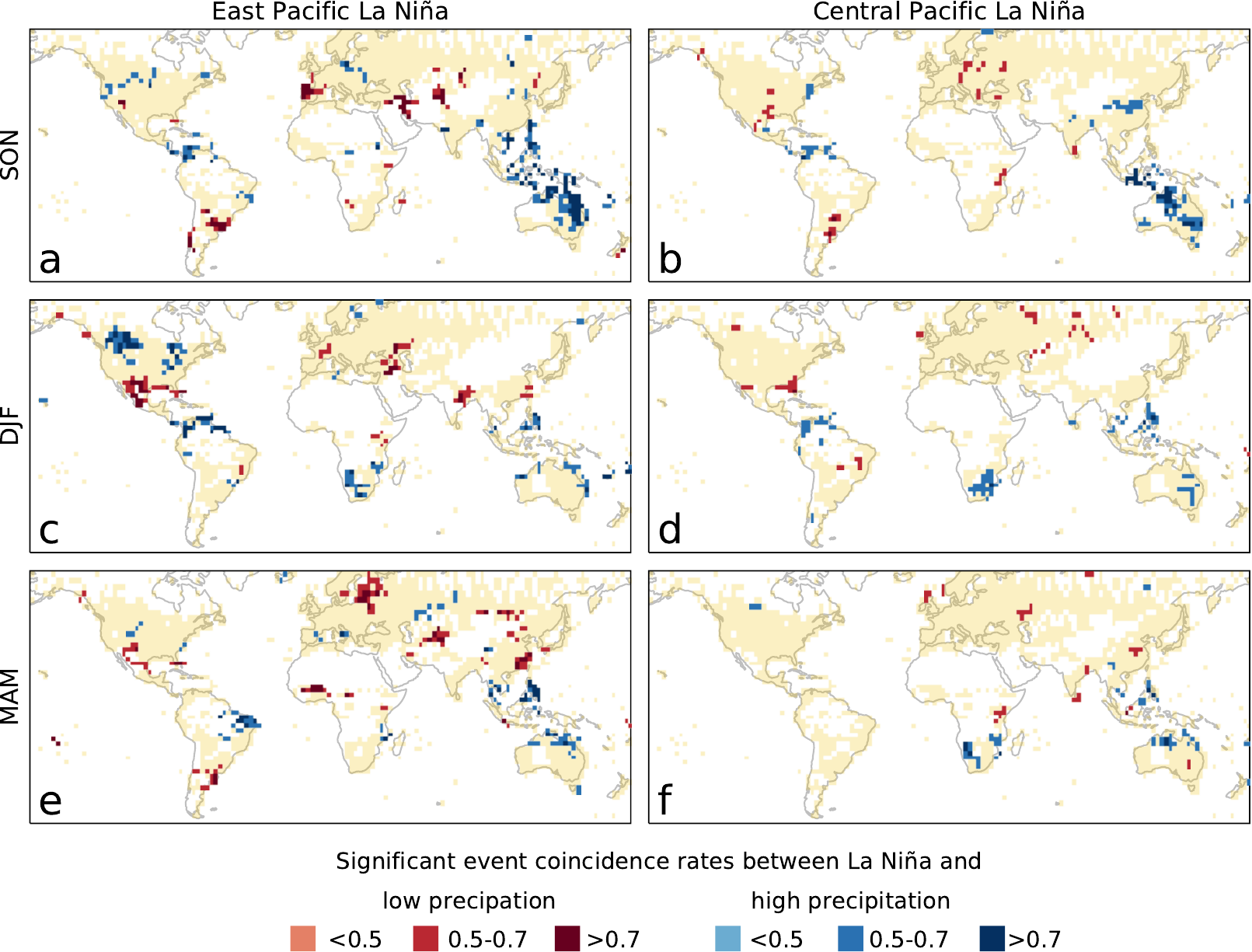}
  \caption{Same as Fig.~\ref{fig:eca_el_nino70} for La Ni\~na periods.}
  \label{fig:eca_la_nina70}
\end{figure*}

To demonstrate the robustness of our results, we repeat the analysis presented in the main manuscript for seasonal precipitation sums falling below/exceeding a threshold defined by the empirical 30/70\% or 10/90\% quantiles, respectively. The corresponding significant event coincidence rates between EP/CP El Ni\~no/La Ni\~na periods and outstanding seasonal precipitation sums above (below) the 90th (10th) percentile are shown in Fig.~\ref{fig:eca_el_nino90} and Fig.~\ref{fig:eca_la_nina90}. Figure~\ref{fig:eca_el_nino70} and~\ref{fig:eca_la_nina70} show the same for strong and weak precipitation defined as values above (below) the 70th (30th) percentile and the respective ENSO periods. The arrangements of the panels corresponds to Fig.\ 2 and\ 3 in the main manuscript. 

We observe that changing the percentile threshold such that only the seasons with the 10\% highest or lowest precipitation sums are considered as events yields an expected reduction in spatially coherent coincidences for both El Ni\~no (Fig.~\ref{fig:eca_el_nino90}) and La Ni\~na periods (Fig.~\ref{fig:eca_la_nina90}). However, the dominant large scale patterns, such as reduced rainfall over Indonesia and the Philippines in SON during EP El Ni\~no phases or strong precipitation over North Australia in SON during EP (and less pronounced also for CP) La Ni\~na phases, persist even for such a stricter definition of precipitation events. Generally, we observe that the patterns displayed in Fig.~\ref{fig:eca_el_nino90} and Fig.~\ref{fig:eca_la_nina90} are also present in Fig.\ 2 and\ 3 of the main manuscript.

Similarly, we observe significant event coincidence rates with the 30\% highest and lowest seasonal precipitation sums being considered as events, Fig.~\ref{fig:eca_el_nino70} and~\ref{fig:eca_la_nina70}. We observe similar patterns as displayed in Fig.\ 2 and\ 3 of the main manuscript, but note that the event coincidence rates (especially for the CP phases of El Ni\~no and La Ni\~na) show a tendency to increase. This is because counting less strong or weak signals leads to an increase in the number of precipitation events as compared to the stricter percentile thresholds chosen in the main manuscript. Since the number of ENSO periods is the same for every choice of precipitation threshold, the event coincidence rate is likely to increase with more precipitation events, thus yielding the increased numbers in Fig.~\ref{fig:eca_el_nino70} and Fig.~\ref{fig:eca_la_nina70}. Generally, we find that the spatial patterns which were observed in Fig.\ 2 and Fig.\ 3 of the main manuscript persist also for less rigid definitions of very high and very low seasonal precipitation. 

In summary, our results generally vary smoothly with the actual choice of percentile thresholds above and below which seasons are considered as an event according to their respective precipitation sum, and we therefore consider the analysis presented in the main manuscript to be sufficiently robust against its actual choice. 

\section{Figures}

\begin{figure*}[t]
  \centering
  \includegraphics[width=.65\linewidth]{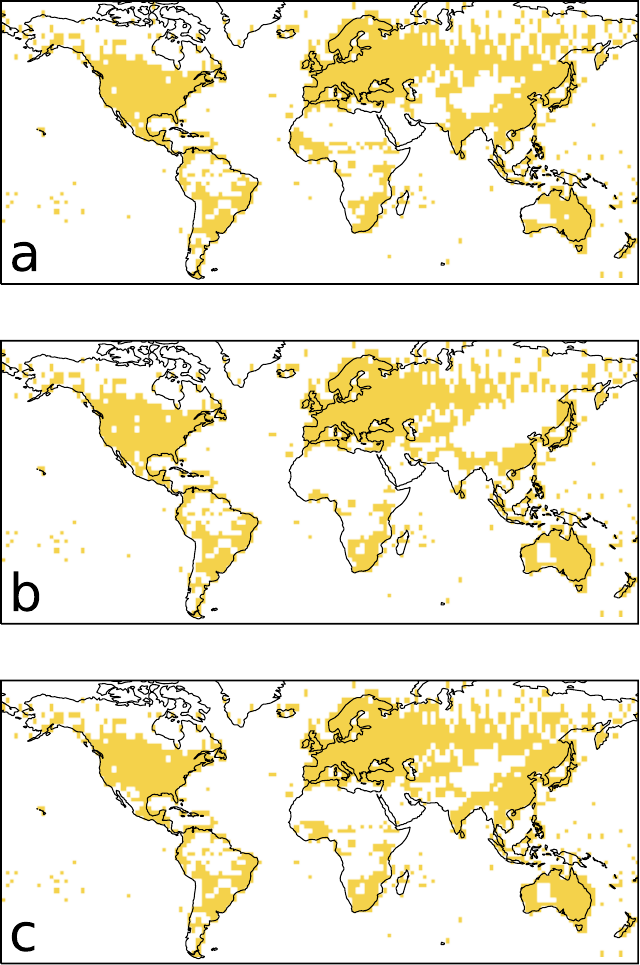}
  \caption{Valid grid cells (yellow) in the GPCC dataset for SON (a), DJF (b) and MAM (c) periods. Grid cells are considered valid if at least one measurement station is present for 95\% of the study period and the average seasonal precipitation sum exceeds 3~cm (see Sec. 2.1 of the main manuscript for details).
  }
\end{figure*}